\documentclass[conference,compsoc]{IEEEtran}

 \usepackage[T1]{fontenc}
 \usepackage[utf8]{inputenc}
 \usepackage{textcomp}
 \usepackage{graphicx}
 \usepackage{multirow}
 \usepackage{xspace}
 \usepackage{listings}
 \usepackage{caption}
 \usepackage{xcolor}
 \usepackage[hyphens]{url}
 \usepackage[hidelinks]{hyperref}
 \usepackage{tikz}
 \usepackage{booktabs}
 \usepackage{amsmath}
 \usepackage{pgf-pie}
 \usepackage{pgfplots}
 \pgfplotsset{compat=1.18}
 
 \usetikzlibrary{positioning,shapes,arrows}

\definecolor{lstbg}{gray}{0.97}
\definecolor{lstkeyword}{rgb}{0.2,0.2,0.7}
\definecolor{lstcomment}{gray}{0.45}
\definecolor{lststring}{rgb}{0.2,0.5,0.3}
\definecolor{lstframe}{gray}{0.75}

\lstdefinestyle{shell}{
  language=bash,
  morekeywords={curl,cat,echo,rm,git,clone},
}

\def\Snospace~{\S{}}

\renewcommand{\figurename}{Fig.}

\newcounter{note}[section]

\newcommand{\appref}[1]{\mbox{Appendix~\ref{#1}}}
\newcommand{\ignore}[1]{}

\providecommand{\attrcirc}[1]{}
\renewcommand{\attrcirc}[1]{%
	\csname tikz\endcsname[baseline=(char.base)]{\node[shape=circle,draw=black,fill=white,line width=0.4pt,inner sep=0.8pt] (char) {\color{black}\scriptsize\bfseries #1};}%
}

\newcommand{\wcirc}{%
	\csname tikz\endcsname[baseline=(char.base)]{%
		\node[shape=circle,draw=black,fill=white,line width=0.4pt,inner sep=0.2ex,minimum size=1.6ex] (char) {\strut};%
	}%
}

\renewcommand{\appref}[1]{\mbox{\autoref{#1}}}

\newcounter{packednmbr}

\definecolor{codebg}{gray}{0.97}
\definecolor{keywordcolor}{rgb}{0.2,0.2,0.7}
\definecolor{commentcolor}{gray}{0.45}
\definecolor{stringcolor}{rgb}{0.2,0.5,0.3}

\lstdefinestyle{rustsec}{
  language=Rust,
  backgroundcolor=\color{codebg},
  basicstyle=\ttfamily\footnotesize,
  keywordstyle=\color{keywordcolor}\bfseries,
  commentstyle=\color{commentcolor}\itshape,
  stringstyle=\color{stringcolor},
  numbers=none,
  numberstyle=\tiny\color{gray},
  breaklines=true,
  breakatwhitespace=true,
  showstringspaces=false,
  columns=fullflexible,
  keepspaces=true,
  captionpos=b,
  frame=single,
  rulecolor=\color{gray!50},
}

\definecolor{codegreen}{rgb}{0,0.6,0}
\definecolor{codegray}{rgb}{0.5,0.5,0.5}
\lstdefinelanguage{Rust}{
    keywords=[1]{as, break, const, continue, crate, else, enum, extern, false, fn, for, if, impl, in, let, loop, match, mod, move, mut, pub, ref, return, self, static, struct, super, trait, true, type, unsafe, use, where, while, async, await, dyn},
    keywords=[2]{Self, Copy, Send, Sized, Sync, Drop, Fn, FnMut, FnOnce, Box, Vec, String, Option, Result, Some, None, Ok, Err, From, Into, Default},
    keywordstyle=[1]\color{codegreen}\bfseries,  
    keywordstyle=[2]\color{blue}\bfseries,  
    commentstyle=\color{codegray},  
    stringstyle=\color{red},  
    morecomment=[l][\color{gray}]{//}, 
    morecomment=[s][\color{gray}]{/*}{*/}, 
    morestring=[b]{"},
}

\lstdefinestyle{lst}{
    language=Rust,
    frame=lines,
    framesep=2mm,
    basicstyle=\ttfamily\footnotesize,
    columns=fixed,
    showstringspaces=false,
    numbersep=3pt,
    escapeinside={||},
    numbers=left,
    stepnumber=1,
    breaklines=true,
    breakatwhitespace=true,
    captionpos=b,
    float=t
}
\usepackage{eso-pic}
\newcommand{\ArxivPublicationNotice}{%
  \AddToShipoutPictureFG*{%
    \AtPageUpperLeft{%
      \put(54,-27){\parbox[t]{504pt}{\fontsize{7.5}{9}\selectfont
      Author manuscript of the paper published in 2026 IEEE Symposium on Security and Privacy Workshops (SPW).\\
      Xu He, Sagar Patel, and Shu Wang. DOI: \href{https://doi.org/10.1109/SPW72489.2026.00047}{10.1109/SPW72489.2026.00047}.\\
      \copyright\ 2026, Xu He. Under license to IEEE.\\
      Post-publication dataset link: \url{https://huggingface.co/datasets/XuViewer/mcp-vulnerability-dataset}.\\
      This maintained dataset extends beyond the frozen 24-record snapshot analyzed in this paper.
      }}%
    }%
  }%
}

\begin{document}
\makeatletter
\gdef\@title{Tracing MCP Security Vulnerabilities: Constructing and Analyzing an MCP Security Vulnerability Dataset}
\makeatother

\author{
\IEEEauthorblockN{Xu He}
\IEEEauthorblockA{\textit{VISA Inc.}\\
xuhe3@visa.com}
\and
\IEEEauthorblockN{Sagar Patel}
\IEEEauthorblockA{\textit{VISA Inc.}\\
sagapate@visa.com}
\and
\IEEEauthorblockN{Shu Wang}
\IEEEauthorblockA{\textit{Palo Alto Networks Inc.}\\
shuwang@paloaltonetworks.com}
}

\maketitle
\ArxivPublicationNotice

\begin{abstract}
As a core protocol for integrating large language models (LLMs) with external data sources and tools, the Model Context Protocol (MCP) has quickly become a key interface in emerging AI agent systems. As the MCP ecosystem expands, reported vulnerabilities have also increased, creating a growing software-security attack surface. However, MCP vulnerability information remains scattered, noisy, and difficult to study systematically. This paper addresses that gap by constructing a curated MCP vulnerability dataset from CVE disclosures. Starting from 119 MCP-related records retrieved through February 2026, we filter unrelated cases out and obtain a validated snapshot of 24 MCP vulnerabilities. Our filtering methodology explicitly addresses acronym ambiguity and achieves 96.7\% inter-rater agreement on a sampled exclusion audit while retaining all known high-severity MCP cases identified in public advisories. Using this initial dataset snapshot, we find that 87.5\% of validated vulnerabilities are Critical or High severity, and that injection and path traversal vulnerabilities account for 66.7\% of cases. We further identify recurring issues involving insufficient input validation, inadequate access control, and protocol-level security gaps. Through six representative case studies, we illustrate common exploitation patterns and map them to practical defenses. We release the dataset, reproducibility tooling, and a sustainable maintenance workflow as a foundation for future MCP security measurement, benchmarking, and defense research.
\end{abstract}


\section{Introduction}

The Model Context Protocol (MCP) has become a fundamental component of modern AI agent systems~\cite{anthropic2024mcp}. MCP provides a standardized way for large language models (LLMs) to interact with external resources such as file systems, databases, and APIs. As a result, an MCP-centric AI tool ecosystem has emerged, including AI-powered programming assistants (e.g., Claude~\cite{anthropic2024claude}, Cursor~\cite{cursor2024}), intelligent agents~\cite{chase2022langchain}, and enterprise AI systems~\cite{openai2023chatgpt}. However, this growing ecosystem also introduces security concerns. Vulnerabilities in MCP implementations can lead to unauthorized data access, remote code execution, and system compromise~\cite{anthropic2024mcp_security}. Despite the growing MCP attack surface, the research community faces a fundamental challenge: the lack of high-quality, curated datasets for systematic security analysis. This gap prevents researchers from understanding vulnerability patterns, developing detection tools, and implementing effective defenses.

Systematic MCP security datasets are scarce for a few reasons. MCP was first released in December 2024~\cite{anthropic2024mcp}. In this short timeframe, the community has been primarily focused on exploring the capabilities and use cases of MCP, rather than its security implications. While news articles and technical reports~\cite{jfrog2025mcp_remote,cisco2026ai_connective_tissue} have occasionally highlighted attacks that exploit MCP to compromise LLM-based applications, existing discussions predominantly center on AI-level risks, such as prompt injection and malicious tool orchestration, rather than the underlying software-level vulnerabilities in MCP implementations themselves. As a result, no curated MCP vulnerability dataset currently exists.

We observe that MCP-related CVEs have been accumulating over recent months, and we take this as an opportunity to conduct a systematic study. We collect and analyze these CVEs to build a curated MCP vulnerability dataset. Given that the acronym "MCP" is shared across unrelated domains such as chip interconnects and Linux kernel drivers, we apply a filtering process to verify that each entry is genuinely related to MCP in the AI context. The resulting release captures a frozen snapshot of 24 validated MCP vulnerabilities disclosed through February 2026. 

We then analyze the collected CVEs and conduct case studies on representative examples. Within this initial snapshot, 87.5\% of validated vulnerabilities are rated Critical or High. The most common vulnerability types are command injection, remote code execution, and path traversal, which together account for the majority of cases. These observations indicate that current MCP implementations often fail at input validation, access control, or both. Through case studies of six high-severity vulnerabilities, we further examine their exploitation paths and identify recurring attack patterns, including insufficient sanitization of tool arguments and overly permissive server configurations. Rather than claiming a comprehensive benchmark, we use these findings to motivate targeted defense recommendations and to show how the dataset can support future automated security auditing research.

We make the following contributions:

\begin{itemize}
\item We construct a curated MCP CVE dataset containing 24 validated vulnerabilities and release the frozen February 2026 snapshot together with validation, normalization, enrichment, and drift-audit tooling.

\item We conduct an empirical analysis of the current MCP disclosure snapshot across severity levels, vulnerability types, affected components, and temporal patterns, providing initial evidence about recurring security issues and likely root causes.

\item We perform in-depth case studies of six representative high-severity vulnerabilities, identifying common exploitation patterns and providing actionable defense recommendations.
\end{itemize}
\section{Related Work}
\label{sec:related}

\subsection{Security Datasets for Software Vulnerabilities}

Software security datasets are crucial for enabling data-driven security research, including vulnerability detection, risk assessment, and defense strategy development.

Projects such as CVEfixes~\cite{bhandari2021cvefixes} and VulDeePecker~\cite{li2018vuldeepecker} extract vulnerability information from CVE databases and link them to source code repositories. These datasets have facilitated research on vulnerability patterns and automated detection. However, these datasets often suffer from three key limitations: (1) \textit{high noise levels} due to incomplete or inconsistent CVE records, (2) \textit{low reusability} caused by domain-specific formats and limited documentation, and (3) \textit{poor sustainability} resulting from lack of regular updates as new vulnerabilities emerge.

Researchers have constructed specialized datasets for specific domains, such as web application vulnerabilities~\cite{alqahtani2020webvuln}, IoT security~\cite{sivanathan2019iot}, and mobile app vulnerabilities~\cite{aafer2015android}. While these datasets address domain-specific needs, they do not cover emerging AI protocols like MCP.

\subsection{MCP Security and Vulnerability Analysis}

Beyond traditional software vulnerabilities, MCP introduces attack vectors tied to its role as an LLM interaction layer. Indirect Prompt Injection (IPI) and Tool Poisoning~\cite{li2026structural} exploit the LLM's processing of external tool outputs: if a tool retrieves malicious content, the model may interpret hidden instructions as legitimate goals, leading to data leakage or privilege escalation. Maloyan and Namiot~\cite{maloyan2026breaking} identified Structural Template Injection, where attackers inject control tokens into retrieved context to cause role confusion and trick the agent into executing unauthorized commands. These attack classes are distinct from the software-level CVEs we study, but they share a common root: insufficient validation of data flowing through MCP's tool interfaces.

Initial security assessments of the MCP ecosystem have highlighted significant risks in model-tool interactions. Radosevich and Halloran performed a safety audit demonstrating that leading models such as Claude 3.7 and Llama 3.3 can be coerced into executing malicious code, facilitating remote access, and exfiltrating credentials through standard MCP servers~\cite{radosevich2025mcp}. Wang et al.~\cite{wang2025mcpguard} introduced MCPGuard, an automated system that identified recurring vulnerability patterns across MCP servers, including path traversal and unauthenticated remote code execution. This work also documented critical CVEs such as CVE-2025-49596 in the official MCP Inspector tool. These efforts focus on individual vulnerabilities or specific attack scenarios; our work complements them by providing a systematic, dataset-driven view of the MCP vulnerability landscape.

\section{MCP Vulnerability Dataset Construction}
\label{sec:dataset}

\subsection{Data Collection}

We collected MCP vulnerabilities from the CVE database~\cite{cve2024}, searching for records matching the keyword "MCP" or "Model Context Protocol" disclosed between January 2025 and February 2026. This returned 119 initial records. After removing duplicates and entries marked as RESERVED or REJECTED, we obtained 108 valid records for further processing.

The main challenge in constructing this dataset is acronym ambiguity. The term "MCP" appears across unrelated domains, semiconductor packaging, Linux kernel drivers, industrial control systems, and healthcare, producing substantial noise in CVE search results. We applied a three-stage filtering process to address this.

In the first stage, we reviewed each record to determine whether "MCP" refers to the Model Context Protocol. Records were retained only if the description explicitly mentioned MCP in an AI or LLM context, such as references to MCP servers, clients, or agents. This excluded 72 records (66.7\%), leaving 36. In the second stage, we checked that each remaining record was associated with a specific MCP component or implementation, such as an MCP server, client, SDK, or MCP-integrated platform. Records that mentioned MCP only in passing were excluded, removing 8 more entries. In the third stage, we excluded records with incomplete information or unconfirmed status, removing 4 additional entries. The final dataset contains 24 validated MCP vulnerabilities, as shown in Table~\ref{tab:paper_labels}.

\begin{figure*}[t]
\centering
\includegraphics[width=\linewidth]{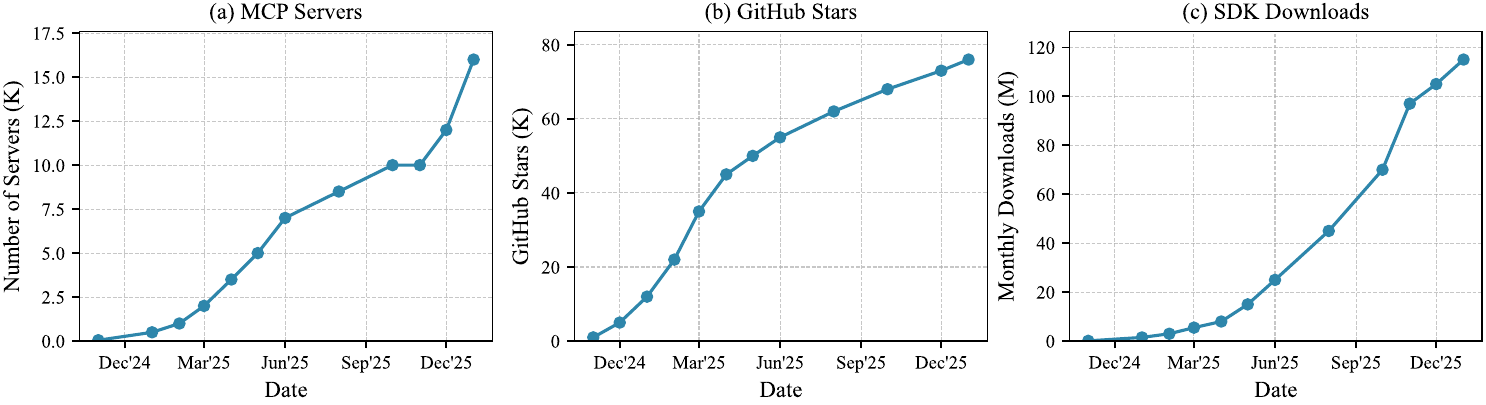}
\caption{Growth of MCP-related repositories on GitHub (2025 to 2026)}
\label{fig:mcp_growth}
\end{figure*}

\subsection{Validation and Availability}

To assess filtering quality, we sampled 30 excluded records and had two independent researchers review them. Both agreed that all sampled exclusions were appropriate (inter-rater agreement: 96.7\%). We also cross-referenced our dataset against Anthropic's security advisories~\cite{anthropic2025security}, vendor reports from JFrog~\cite{jfrog2025mcp} and Trend Micro~\cite{trend2025mcp}, and GitHub Security Advisories for MCP repositories. This validation does not claim that the dataset is exhaustive; rather, it shows that the released snapshot is a reliable starting point for studying publicly disclosed MCP vulnerabilities while minimizing obvious false positives caused by acronym ambiguity. The manuscript results reported in this paper use a frozen 24-record snapshot covering disclosures through February 2026.

The released repository distinguishes between the raw seed CSV, the normalized analysis CSV, the curated paper-label file, and the optional enrichment output. Paper-facing counts are generated from the normalized dataset plus the annotation layer, and the resulting \texttt{paper\_stats.json} file is the authoritative machine-readable summary for the statistics reported in Section~\ref{sec:analysis}. The current repository does not claim a fully automated crawler; instead, it provides reproducible validation, normalization, enrichment, and drift-audit tooling around this frozen snapshot. For transparency, \appref{app:paper_labels} lists the curated paper-label annotation layer used by the analysis.

\begin{figure}[t]
\centering
\includegraphics[width=0.7\linewidth]{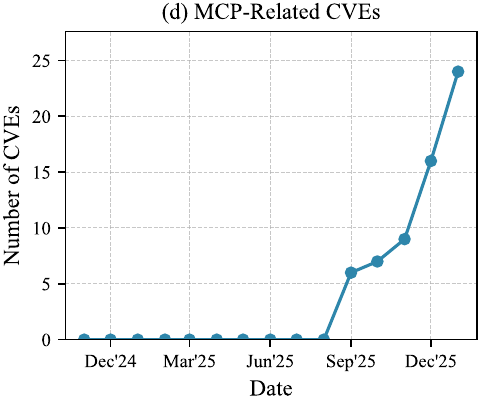}
\caption{Temporal Distribution of MCP Vulnerabilities in the Released Snapshot}
\label{fig:temporal_dist}
\end{figure}

\section{Multi-Dimensional Empirical Analysis}
\label{sec:analysis}

Based on our curated dataset of 24 validated MCP vulnerabilities, this section presents an empirical analysis across four key dimensions: severity distribution, vulnerability types, affected components, and temporal patterns. We also analyze the root causes of MCP vulnerabilities and their correlations.

To provide context for the dataset, Figure~\ref{fig:mcp_growth} shows the growth of MCP-related repositories on GitHub from early 2025 to early 2026. The rapid increase in project count reflects the fast-paced adoption of MCP across the developer community, which in turn expands the attack surface and motivates systematic vulnerability tracking.

\subsection{Severity and Temporal Distribution}

Among the 24 vulnerabilities, 14 are rated Critical (58.3\%), 7 High (29.2\%), 2 Medium (8.3\%), and 1 Low (4.2\%). The resulting 87.5\% Critical/High rate is high relative to typical CVE distributions across all software~\cite{cvss2023stats}, although this comparison should be interpreted cautiously given the small sample size and disclosure-driven nature of our dataset. One plausible explanation is that MCP servers often hold privileged access to file systems, databases, and APIs, which can amplify the impact of implementation flaws.

As shown in Figure~\ref{fig:temporal_dist}, the released snapshot is temporally uneven. Fourteen records fall in January 2025 because many source records only disclose the year, and our normalization pipeline maps year-only dates to \texttt{YYYY-01-01} while preserving a separate date-precision flag. Beyond that normalization artifact, the next clear burst appears in January 2026 with 7 entries, following smaller increases in September 2025 (1) and December 2025 (2). The overall pattern still suggests that vulnerability discovery in the MCP ecosystem is in an early and rapidly changing stage, but month-level conclusions should be interpreted cautiously because most records do not yet provide exact publication dates.

\subsection{Vulnerability Types and Affected Components}

Table~\ref{tab:vuln_types} shows the distribution of vulnerability types across the 24 entries. Command injection (7), path traversal (5), and remote code execution (4) together account for 66.7\% of all cases. DNS rebinding (2), unauthorized access (2), hardcoded credentials (1), ReDoS (1), XSS (1), and one residual Other category make up the remainder.

\begin{table}[t]
\centering
\caption{MCP Vulnerability Type Distribution}
\label{tab:vuln_types}
\begin{tabular}{p{3.5cm}cp{2.5cm}}
\toprule
\textbf{Vulnerability Type} & \textbf{Count} & \textbf{Primary CWE} \\
\midrule
Command Injection & 7 & CWE-77, CWE-78 \\
Remote Code Execution & 4 & CWE-94 \\
Path Traversal & 5 & CWE-22 \\
Unauthorized Access & 2 & CWE-306, CWE-862 \\
ReDoS & 1 & CWE-1333 \\
XSS & 1 & CWE-79 \\
DNS Rebinding & 2 & CWE-346 \\
Hardcoded Credentials & 1 & CWE-798 \\
Other & 1 & N/A \\
\bottomrule
\end{tabular}
\end{table}

In terms of affected components, 11 vulnerabilities (45.8\%) are in MCP servers, 8 (33.3\%) in MCP clients, and 5 (20.8\%) in MCP SDKs. Looking at frequently affected components, Anthropic's \texttt{mcp-server-git} contributes three Critical entries (CVE-2025-68143, CVE-2025-68144, and CVE-2025-68145); Orval contributes two client-side entries (CVE-2026-22785 and CVE-2026-23947); and the MCP TypeScript SDK contributes two widely propagated issues (CVE-2025-66414 and CVE-2026-0621).

Examining the relationship between vulnerability types and affected components reveals a meaningful distinction. The majority of server-side vulnerabilities are command injection, RCE, and path traversal. These are classic software-level flaws caused by insufficient input validation when MCP servers pass client-supplied arguments to underlying system calls or file operations. They are bugs in the MCP \emph{implementation} itself: the server accepts a parameter, fails to sanitize it, and exposes the host system. \texttt{mcp-server-git} is a clear example, where unsanitized git parameters lead directly to command injection.

Client-side and SDK vulnerabilities, by contrast, tend to involve unauthorized access, hardcoded credentials, DNS rebinding, and XSS. These flaws arise not in how MCP processes tool arguments, but in how AI clients authenticate with MCP servers, how they handle server-returned content, or how the integration layer is configured. For instance, DNS rebinding in the TypeScript SDK (CVE-2025-66414) allows a malicious web page to reach a locally running MCP server by exploiting the HTTP transport layer, an issue that manifests at the boundary between the AI application and the MCP server, rather than within the server's core logic.

This distinction echoes a point raised in the introduction: prior security discussions around MCP have focused on AI-level threats such as prompt injection and tool misuse, while software-level flaws in MCP implementations have received less attention. Our dataset shows that both dimensions are present in current public disclosures. Server-side implementation bugs are more numerous and directly exploitable without any LLM involvement. Client-side and integration flaws, while fewer, represent the attack surface that emerges specifically from deploying MCP within AI application stacks.

\subsection{Root Cause Analysis}

The vulnerability type and component analysis in the previous section points to three overlapping root-cause categories rather than mutually exclusive buckets.
Appendix~\ref{app:paper_labels} provides the per-CVE annotation layer underlying this analysis, including the vulnerability-type, component, and root-cause labels for all 24 cases.
The dominant root cause is insufficient input validation. 18 out of 24 vulnerabilities (75.0\%) involve MCP implementations passing client-supplied inputs directly to system-level operations, file-path resolvers, code generators, or parsers without adequate sanitization. This is not coincidental: MCP's design encourages servers and client tooling to expose powerful operations as callable tools, which creates many opportunities for untrusted inputs to reach sensitive sinks. CVE-2025-68144 and CVE-2026-22785 are representative examples in which attacker-controlled parameters propagate into command execution paths in \texttt{mcp-server-git} and Orval.

The second category is inadequate access control, present in 4 vulnerabilities (16.7\%). These cases include insufficient authorization around command execution in Flowise (CVE-2025-8943), hardcoded shared secrets in Arcade MCP (CVE-2025-66454), deeplink-driven configuration tampering in Dive (CVE-2026-23523), and overly permissive command execution in LibreChat (CVE-2026-22252). Although fewer in number than the input-validation cases, they show the same pattern of security being deferred in favor of frictionless setup or local convenience.

The third category is protocol-level security gaps, present in 3 vulnerabilities (12.5\%) and sometimes overlapping with the previous two categories. The initial MCP specification defined message formats and interaction patterns for functional interoperability, but left several transport and parser-hardening requirements unspecified. The two DNS rebinding entries (CVE-2025-66414 and CVE-2025-9611) and the UriTemplate ReDoS in the TypeScript SDK (CVE-2026-0621) illustrate how these omissions propagate across both SDKs and downstream applications.

In short, the root causes form a chain: specification gaps appear to have left room for unsafe defaults, and those defaults repeatedly manifested as input-validation failures, access-control mistakes, or both in the current MCP disclosure snapshot.

\subsection{Summary}

Taken together, the analysis paints a consistent picture within the current dataset snapshot. MCP vulnerabilities are predominantly severe, concentrated in input-validation failures, and distributed across both server implementations and client/SDK integrations. The current disclosure pattern suggests that the ecosystem is still at an early security maturity stage, but larger and longer-term datasets are needed before drawing ecosystem-wide conclusions. At minimum, the released snapshot indicates that improving MCP security will require both implementation-level fixes and clearer protocol-level guidance.

\begin{table*}[t]
\centering
\caption{Summary of Representative Vulnerability Case Studies}
\label{tab:case_summary}
\begin{tabular}{p{2.2cm}p{1.6cm}p{2.5cm}p{4.5cm}p{4.5cm}}
\toprule
\textbf{CVE ID} & \textbf{Component} & \textbf{Type} & \textbf{Root Cause} & \textbf{Key Fix} \\
\midrule
CVE-2025-68144 & Server & Command Injection & Unsanitized shell execution & Use library APIs; allowlist validation \\
CVE-2025-68145 & Server & Path Traversal & Incomplete path normalization & Canonicalize then check containment \\
CVE-2026-22785 & Client & Command Injection & Unescaped OpenAPI metadata & Escape inputs; parameterized execution \\
CVE-2026-23523 & Client & Unauthorized Access & Insufficient confirmation on deeplink installs & Require explicit approval for server installs \\
CVE-2025-66414 & SDK & DNS Rebinding & No rebinding protection in HTTP transport & Enable DNS rebinding protection in transport layer \\
CVE-2026-0621  & SDK & ReDoS & Catastrophic backtracking in UriTemplate processing & Harden UriTemplate parsing for RFC 6570 patterns \\
\bottomrule
\end{tabular}
\end{table*}

\section{Representative Vulnerability Case Studies}
\label{sec:cases}

We examine six representative high-severity vulnerabilities selected to cover the two component classes and three root causes identified in Section~\ref{sec:analysis}. Table~\ref{tab:case_summary} provides an overview. For each case, we describe the exploitation path, the applied fix, and the broader takeaway.

\subsection{Server-Side Cases}

\noindent\textbf{Case 1: Command Injection in mcp-server-git (CVE-2025-68144)~\cite{CVE-2025-68144}.}
\texttt{mcp-server-git} is Anthropic's official MCP server for git operations. Prior to version 2025.12.17, user-controlled arguments in tools such as \texttt{git\_diff} and \texttt{git\_checkout} were passed directly to Git CLI commands without sanitization. Because MCP clients supply tool arguments as structured JSON that is then forwarded into command execution paths, an attacker who controls a client, or who can influence tool arguments through prompt injection, can inject arbitrary shell behavior. As shown in the following box, an example payload is to smuggle an attacker-controlled flag value such as \texttt{--output} into the downstream command invocation.

\begin{lstlisting}[style=shell]
git_diff_args = ["--output=; curl attacker.com/steal?d=$(cat /etc/passwd)"]
\end{lstlisting}

When the server executed the resulting Git command, the injected payload ran with the privileges of the server process, which typically has access to the developer's local file system and credentials. This vulnerability is rated Critical. Anthropic fixed it in version 2025.12.17 by sanitizing the affected arguments and tightening how user-controlled values are forwarded into Git execution paths. The key lesson is that MCP servers frequently wrap system utilities to expose them as tools; any such wrapper must avoid shell interpolation entirely and prefer library APIs or subprocess interfaces that do not invoke a shell.

\noindent\textbf{Case 2: Path Traversal in mcp-server-git (CVE-2025-68145)~\cite{CVE-2025-68145}.}
The same server also exposed repository-scoped operations intended to remain inside a path supplied via the \texttt{--repository} flag. Before version 2025.12.17, the server did not correctly validate \texttt{repo\_path} arguments against that repository restriction. Standard traversal-style inputs or alternate path encodings could therefore steer the server toward unintended repositories or filesystem locations:

\begin{lstlisting}[style=shell]
repo_path = "../another-repo/.git"
\end{lstlisting}

Because the repository-boundary check was incomplete, an attacker could escape the intended repository sandbox and operate on content outside the trusted path. The vulnerability was marked as Critical. Version 2025.12.17 fixed this by validating repository paths against the configured repository restriction before carrying out the underlying operation. This case illustrates a common class of validation bypass: access-control checks over filesystem inputs must operate on the same normalized path representation that the underlying tool will eventually use.

\subsection{Client-Side Cases}

\noindent\textbf{Case 3: Command Injection in Orval (CVE-2026-22785)~\cite{CVE-2026-22785}.}
Orval is a widely-used code generator that can produce MCP client code from OpenAPI specifications. Before version 7.18.0, its MCP server generation logic relied on string manipulation that incorporated unsanitized \texttt{summary} fields from the OpenAPI specification. A developer who pointed Orval at a malicious or compromised specification could therefore trigger attacker-controlled code execution during code generation:

\begin{lstlisting}[style=shell]
summary: "endpoint\"; require('child_process').exec('curl attacker'); //"
\end{lstlisting}

By breaking the generated string literal, the attacker can splice arbitrary code into the generation flow and compromise the developer's build environment. The attack surface is particularly broad because OpenAPI specifications are routinely fetched from third-party services, public registries, or community repositories. Version 7.18.0 fixed this by sanitizing the affected metadata field before it is incorporated into generated output. This case is notable for two reasons: first, the vulnerable code path is triggered by static data processing rather than interactive runtime input, which makes it easy to overlook in security reviews; second, it illustrates a supply chain risk specific to the MCP ecosystem, where client tooling processes external API definitions as part of normal development workflow.

\noindent\textbf{Case 4: Unauthorized Configuration Tampering in Dive (CVE-2026-23523)~\cite{CVE-2026-23523}.}
Dive is a desktop MCP host that manages connections to multiple MCP servers and their associated permissions. Prior to version 0.13.0, crafted deeplinks could install attacker-controlled MCP server configurations without sufficient user confirmation. That meant a user could be lured into opening a malicious link which then added a new MCP server, or replaced an existing one, with too little friction for the security sensitivity of the action. Because Dive acts as the trust broker between the AI application and all connected MCP servers, compromising the configuration layer effectively compromises every tool the AI application can invoke. Version 0.13.0 fixed this by adding stronger confirmation requirements before new server configurations can be installed through deeplinks. This case highlights that the security boundary in an MCP deployment is not just the server or the AI model, but also the client application that decides which servers become trusted.

\subsection{SDK Cases}

\noindent\textbf{Case 5: DNS Rebinding in MCP TypeScript SDK (CVE-2025-66414)~\cite{CVE-2025-66414}.}
The MCP TypeScript SDK's HTTP transport layer originally shipped without DNS rebinding protection enabled. DNS rebinding exploits this by using an attacker-controlled domain with a very short TTL. In the first phase, the domain resolves to an attacker-approved external IP, allowing an initial connection or trust decision to succeed. After the TTL expires, the attacker updates the DNS record to point to an internal address such as \texttt{127.0.0.1} or a private network range. When the client reconnects or reuses the hostname, it now reaches an unintended internal service, effectively treating the MCP client as a proxy into the local network. Any MCP server or local HTTP service reachable from the client machine becomes accessible to the attacker. The vulnerability was assigned a High severity rating because MCP clients commonly run on developer workstations alongside other local services. Version 1.24.0 mitigated this by enabling DNS rebinding protection in the transport layer. Because the TypeScript SDK is a foundational shared dependency used across many downstream projects, the fix propagated broadly once adopters updated their dependency.

\noindent\textbf{Case 6: ReDoS in MCP TypeScript SDK (CVE-2026-0621)~\cite{CVE-2026-0621}.}
Versions of the MCP TypeScript SDK up to and including 1.25.1 contained a regular-expression denial-of-service vulnerability in the \texttt{UriTemplate} class when processing RFC 6570 exploded array patterns. The affected pattern could trigger catastrophic backtracking in backtracking-based regex engines. By supplying a carefully crafted template input that repeatedly forces partial matches to fail late, an attacker can drive the engine into extremely expensive evaluation and exhaust CPU resources. In an MCP deployment that expands URI templates while servicing requests, this can stall the server long enough to constitute a denial of service. The fix shipped in version 1.25.1 and hardened \texttt{UriTemplate} processing for the affected RFC 6570 pattern. This case is a reminder that ReDoS is not merely a theoretical concern: in protocol implementations where structured strings are parsed from partially trusted sources, parser complexity must be treated as part of the attack surface.

\subsection{Cross-Case Discussion}

Across all six cases, several consistent patterns emerge that align with the root cause analysis in Section~\ref{sec:analysis}.

First, server-side bugs (Cases 1 and 2) share the same underlying structure: the server receives a parameter from an MCP client, performs either no validation or a flawed validation, and passes the value to a system-level operation. In both cases the fix required correcting the validation logic, either moving it to the right stage (after normalization) or eliminating shell interpolation entirely. These are classic software implementation flaws that appear regardless of MCP; what MCP adds is a standardized pathway for external inputs to reach these operations from AI clients.

Second, client-side bugs (Cases 3 and 4) arise at the integration boundary rather than inside the core MCP protocol logic. Case 3 is unusual in that the vulnerable path is triggered by static data processing (code generation), not runtime execution; Case 4 shows that the client application itself can be a privileged attack target because it controls the trust relationships between the AI model and all connected servers. Both cases illustrate that securing an MCP deployment requires attention to the entire toolchain, not just the server implementations.

Third, SDK-level bugs (Cases 5 and 6) have the widest blast radius. A single vulnerability in the TypeScript SDK propagates to every project that depends on it, and the affected projects may never have written the vulnerable code themselves. This creates an asymmetric incentive: effort spent fixing security issues at the SDK level yields broad ecosystem benefits at low marginal cost, whereas leaving such protections to individual applications means the same vulnerability class will recur independently across many projects.

Finally, none of the six vulnerabilities required novel attack techniques. Command injection, path traversal, unauthorized configuration tampering, DNS rebinding, and ReDoS are all well-understood vulnerability classes with established mitigations. Their presence in the MCP ecosystem at this concentration, and in official and widely adopted implementations, reflects the absence of security requirements in the MCP protocol specification rather than isolated developer error. Addressing that gap at the protocol level is the most scalable path to reducing the vulnerability density observed in this dataset.

\section{Defense Strategies and Dataset Sustainability}
\label{sec:defense}

The root cause analysis in Section~\ref{sec:analysis} identified insufficient input validation as the dominant failure pattern, with recurring access-control mistakes and protocol-level security gaps as secondary contributors. This section translates those findings into concrete defense recommendations, organized by the layer at which each countermeasure applies. We then describe the maintenance strategy for keeping the dataset current.

\subsection{Defense Recommendations}

\noindent\textbf{Protocol specification hardening.}
The most scalable intervention is to close the security gaps in the MCP specification itself. Currently, the specification defines message formats and interaction patterns but leaves input validation, authentication, and transport protection largely unspecified. This allows independent implementations to make inconsistent security decisions, which is why the same vulnerability classes recur across unrelated projects. Protocol maintainers should add mandatory requirements for input validation (allowlist-based parameter schemas, maximum input lengths, character set constraints), authentication (OAuth or mTLS for networked transports, API token authentication for local HTTP endpoints), and transport security (TLS for network transports, DNS rebinding protection for HTTP transports, per-session IP pinning). Secure defaults such as deny-by-default tool permissions and mandatory audit logging should be part of the baseline rather than optional recommendations.

\noindent\textbf{Secure implementation practices.}
For server developers, the central rule is never to pass client-supplied inputs directly to system operations. Shell commands should be replaced by library API calls that accept structured arguments; file paths should be canonicalized before containment checks are applied; SQL queries should use parameterized interfaces; and regex patterns that validate structured protocol messages should be replaced by purpose-built parsers. These practices directly address the command injection, path traversal, and remote code execution patterns that together account for 66.7\% of the dataset. Servers should also run under the least privilege required: non-root accounts, filesystem access restricted to the working directory, and outbound network access limited to what the tool actually needs. Credentials must never be hardcoded; environment variables or secrets managers should be used instead, as CVE-2025-66454 illustrates the consequences when this rule is ignored.

\noindent\textbf{Client and integration security.}
Client developers and MCP application integrators face a different threat model. The case studies in Section~\ref{sec:cases} show that client-side vulnerabilities tend to involve insufficient confirmation around trust-establishing actions (CVE-2026-23523) and unescaped external data processed during code generation (CVE-2026-22785). Configuration changes triggered by deeplinks, imports, or background discovery should require explicit user approval, because those actions effectively decide which MCP servers become trusted. External data such as OpenAPI specifications, server-returned metadata, and tool descriptions must be treated as untrusted and escaped or sanitized before use in commands or code generation pipelines.

\noindent\textbf{Dependency management and patching.}
Several vulnerabilities in the dataset, including the two SDK cases, affect shared dependencies that propagate to many downstream projects. Operators should maintain an inventory of all MCP components and their versions, with automated dependency scanning in CI/CD pipelines to surface newly disclosed CVEs. High-priority patches identified in this study include \texttt{mcp-server-git} ($\geq$\,2025.12.17), Orval ($\geq$\,8.0.2 to cover both Orval entries in our dataset), and the MCP TypeScript SDK ($\geq$\,1.25.1). Unused high-risk tools such as file deletion, repository initialization, and arbitrary command execution should be disabled rather than left exposed.

\noindent\textbf{Runtime monitoring.}
Even well-implemented servers benefit from runtime controls. MCP servers should log all tool invocations with their full parameter values; these logs make injection attempts visible and support post-incident forensics. Anomaly detection rules keyed on shell metacharacters in parameters, \texttt{../} sequences in path arguments, and unusually large input payloads can catch exploitation attempts before they succeed. Servers handling sensitive operations should run in containers or VMs to limit the blast radius of a successful exploit. For HTTP-based transports, rate limiting and Host header validation reduce the attack surface for both DoS and DNS rebinding attacks.

\subsection{Dataset Sustainability}

A vulnerability dataset for a rapidly evolving ecosystem requires an explicit maintenance strategy to remain useful. The MCP ecosystem is still in an early security maturity stage; disclosure rates are accelerating, as shown in Section~\ref{sec:analysis}, so the current 24-entry dataset is a snapshot rather than a complete picture.

The current repository does not implement a fully automated crawler. Instead, it provides reproducible tooling for validation, normalization, enrichment, and manuscript-drift auditing around the frozen February 2026 snapshot used in this paper. New candidate CVEs are still identified through periodic review of NVD, Anthropic's official security advisories, GitHub Security Advisories for MCP repositories, and vendor reports such as JFrog and Trend Micro. This workflow reduces the risk of missed disclosures while keeping analyst judgment explicit when deciding whether a record truly belongs in the MCP dataset.

Future dataset versions can add fields that are not yet consistently available, such as precise affected version ranges, links to patch commits, PoC exploit references where publicly disclosed, and exploitation difficulty ratings. The dataset is released in CSV and JSON formats under CC BY 4.0 and hosted on GitHub with the filtering scripts and documentation. We welcome community contributions: reports of missing entries, corrections to existing records, and improvements to the validation and enrichment workflow.

\section{Discussion}
\label{sec:discussion}

\subsection{Limitations}

Our dataset covers CVE records disclosed through February 2026 and is necessarily incomplete: vulnerabilities discovered and fixed without public disclosure are not captured, and new entries will emerge frequently as the MCP ecosystem continues to grow. The 18-month temporal window since MCP's release (late 2024) is too short to draw conclusions about long-term vulnerability lifecycle patterns or ecosystem maturation trends; those questions require multi-year follow-up studies. In addition, 20 of the 24 records currently expose only year-level disclosure dates in the released sources, so our temporal analysis should be interpreted as a coarse trend view rather than a precise month-by-month disclosure history.

The filtering process involves inherent subjectivity, particularly in the component association stage where determining whether a CVE targets an MCP-specific component requires judgment calls that automated keyword matching cannot fully replace. Despite a 96.7\% inter-rater agreement in our validation sample, some borderline cases may have been classified inconsistently. We also did not independently verify that disclosed patches fully remediate the reported vulnerabilities or that no regressions were introduced; our case study analysis relies on vendor-provided fix descriptions.

The small size of the current dataset also constrains what can be concluded from it. With only 24 validated vulnerabilities drawn from public disclosures in a young ecosystem, the dataset is better understood as an initial research resource than as a mature benchmark for downstream learning or detection tasks. Our empirical analysis therefore aims to characterize an early snapshot of publicly visible MCP security issues, not to claim a comprehensive picture of all MCP vulnerabilities.

\subsection{Future Work}

Three directions stand out as promising follow-ons to this work.

\noindent\textbf{Automated vulnerability detection.}
The dataset provides labeled examples of injection, path traversal, and access control vulnerabilities in MCP server and client code. These could be used to train static analysis models or fine-tune code-understanding models to flag similar patterns proactively, before vulnerabilities are disclosed. Data flow analysis of how MCP tool arguments propagate to system calls would be a natural starting point, as this pattern underlies the majority of cases in the dataset.

\noindent\textbf{Cross-protocol comparison.}
MCP is one of several protocols that expose AI model capabilities to external tools, alongside OpenAI Function Calling, LangChain tool integration, and Semantic Kernel. It is not yet known whether the vulnerability distribution observed here, with heavy concentration in injection and access control driven by protocol specification gaps, is specific to MCP's design or a general property of AI tool-use protocols. A comparative dataset spanning multiple protocols would clarify which design choices correlate with better security outcomes and whether unified security requirements are feasible.

\noindent\textbf{Protocol specification improvement.}
The root cause analysis points to a concrete, actionable research direction: drafting and evaluating formal security requirements for the MCP specification. This would involve proposing specific normative language for input validation, authentication, and transport security, assessing how existing implementations would need to change to comply, and measuring the impact on vulnerability rates over time as the specification evolves.

\section{Conclusion}
\label{sec:conclusion}

We presented the first curated dataset of MCP vulnerabilities, constructed from 119 CVE records through a three-stage filtering pipeline designed to handle the acronym ambiguity that makes naive keyword searches unreliable for this protocol. The released February 2026 snapshot contains 24 validated MCP vulnerabilities and provides an initial basis for studying software-security issues in the MCP ecosystem. Our analysis shows that most currently disclosed cases are Critical or High severity, that command injection, path traversal, and remote code execution account for most entries, and that server implementations, clients, and SDK layers are all affected. The six case studies further illustrate recurring exploitation patterns and connect them to practical defenses. Rather than claiming a complete view of MCP security, we release a reproducible, extensible dataset and tooling stack that future work can build on for longitudinal measurement, benchmarking, and defense evaluation.

\appendices
\section{Reproducibility and Annotation Transparency}
\label{app:paper_labels}

This appendix (Table \ref{tab:paper_labels}) documents indexs of 24 CVEs analyzed in this paper. Each record is associated with a vulnerability-type label, a coarse component bucket, and three non-exclusive root-cause flags, which are corresponding to the analysis results in Section \ref{sec:analysis} and \ref{sec:cases}. The flags indicate whether the case involves insufficient input validation (IV), inadequate access control (AC), or a protocol-level security gap (PG).

\begin{table}[t]
\centering
\scriptsize
\caption{Details of 24 analyzed CVE vulnerabilities.}
\label{tab:paper_labels}
\begin{tabular}{llcccc}
\toprule
\textbf{CVE ID} & \textbf{Vulnerability Type} & \textbf{Comp.} & \textbf{IV} & \textbf{AC} & \textbf{PG} \\
\midrule
CVE-2026-23947 & Remote Code Execution & client & Y & N & N \\
CVE-2026-23744 & Remote Code Execution & server & Y & N & N \\
CVE-2026-23523 & Unauthorized Access & client & N & Y & N \\
CVE-2026-22785 & Command Injection & client & Y & N & N \\
CVE-2026-22688 & Command Injection & sdk & Y & N & N \\
CVE-2026-22252 & Command Injection & client & Y & Y & N \\
CVE-2026-0621 & ReDoS & sdk & Y & N & Y \\
CVE-2025-9654 & Command Injection & server & Y & N & N \\
CVE-2025-9611 & DNS Rebinding & server & N & N & Y \\
CVE-2025-9262 & Command Injection & client & Y & N & N \\
CVE-2025-8943 & Unauthorized Access & server & N & Y & N \\
CVE-2025-69256 & Command Injection & sdk & Y & N & N \\

CVE-2025-68433 & Remote Code Execution & client & Y & N & N \\
CVE-2025-68205 & Other & server & N & N & N \\
CVE-2025-68145 & Path Traversal & server & Y & N & N \\
CVE-2025-68144 & Command Injection & server & Y & N & N \\
CVE-2025-68143 & Path Traversal & server & Y & N & N \\
CVE-2025-67366 & Path Traversal & server & Y & N & N \\
CVE-2025-67364 & Path Traversal & server & Y & N & N \\
CVE-2025-66689 & Path Traversal & server & Y & N & N \\
CVE-2025-66580 & XSS & client & Y & N & N \\
CVE-2025-66562 & Remote Code Execution & client & Y & N & N \\
CVE-2025-66454 & Hardcoded Credentials & sdk & N & Y & N \\
CVE-2025-66414 & DNS Rebinding & sdk & N & N & Y \\
\bottomrule
\end{tabular}
\end{table}

The table mirrors the paper-facing annotation CSV used by the analysis pipeline, allowing readers to inspect the coding decisions directly without consulting the released repository.

\bibliographystyle{IEEEtran}
\bibliography{ref}

\begin{thebibliography}{10}
\providecommand{\url}[1]{#1}
\csname url@samestyle\endcsname
\providecommand{\newblock}{\relax}
\providecommand{\bibinfo}[2]{#2}
\providecommand{\BIBentrySTDinterwordspacing}{\spaceskip=0pt\relax}
\providecommand{\BIBentryALTinterwordstretchfactor}{4}
\providecommand{\BIBentryALTinterwordspacing}{\spaceskip=\fontdimen2\font plus
\BIBentryALTinterwordstretchfactor\fontdimen3\font minus
  \fontdimen4\font\relax}
\providecommand{\BIBforeignlanguage}[2]{{%
\expandafter\ifx\csname l@#1\endcsname\relax
\typeout{** WARNING: IEEEtran.bst: No hyphenation pattern has been}%
\typeout{** loaded for the language `#1'. Using the pattern for}%
\typeout{** the default language instead.}%
\else
\language=\csname l@#1\endcsname
\fi
#2}}
\providecommand{\BIBdecl}{\relax}
\BIBdecl

\bibitem{anthropic2024mcp}
{Anthropic}, ``{Model Context Protocol (MCP)},''
  \url{https://modelcontextprotocol.io}, 2024, accessed: 2026-02-27.

\bibitem{anthropic2024claude}
\BIBentryALTinterwordspacing
Anthropic, ``Claude,'' 2024, accessed: 2026-04-07. [Online]. Available:
  \url{https://www.anthropic.com/claude}
\BIBentrySTDinterwordspacing

\bibitem{cursor2024}
\BIBentryALTinterwordspacing
Anysphere, ``Cursor - the ai code editor,'' 2024, accessed: 2026-04-07.
  [Online]. Available: \url{https://www.cursor.com/}
\BIBentrySTDinterwordspacing

\bibitem{chase2022langchain}
H.~Chase, ``Langchain,'' \url{https://github.com/langchain-ai/langchain}, 2022,
  accessed: 2026-04-07.

\bibitem{openai2023chatgpt}
\BIBentryALTinterwordspacing
OpenAI, ``Chatgpt,'' 2023, accessed: 2026-04-07. [Online]. Available:
  \url{https://openai.com/chatgpt}
\BIBentrySTDinterwordspacing

\bibitem{anthropic2024mcp_security}
\BIBentryALTinterwordspacing
Anthropic, ``Model context protocol (mcp) security best practices,'' 2024,
  accessed: 2026-04-07. [Online]. Available:
  \url{https://modelcontextprotocol.io/docs/tutorials/security/security_best_practices}
\BIBentrySTDinterwordspacing

\bibitem{jfrog2025mcp_remote}
\BIBentryALTinterwordspacing
J.~S. Research, ``Critical {RCE} vulnerability in mcp-remote: {CVE-2025-6514}
  threatens {LLM} clients,'' July 2025, accessed: 2026-04-07. [Online].
  Available:
  \url{https://jfrog.com/blog/2025-6514-critical-mcp-remote-rce-vulnerability/}
\BIBentrySTDinterwordspacing

\bibitem{cisco2026ai_connective_tissue}
\BIBentryALTinterwordspacing
C.~Systems, ``Ai's `connective tissue' is woefully insecure,''
  \emph{Cybersecurity Dive}, February 2026, accessed: 2026-04-07. [Online].
  Available:
  \url{https://www.cybersecuritydive.com/news/ai-agents-model-context-protocol-cisco-report/812580/}
\BIBentrySTDinterwordspacing

\bibitem{bhandari2021cvefixes}
G.~Bhandari, A.~Naseer, and L.~Moonen, ``{CVEfixes: Automated Collection of
  Vulnerabilities and Their Fixes from Open-Source Software},'' in
  \emph{Proceedings of the 17th International Conference on Predictive Models
  and Data Analytics in Software Engineering}, 2021, pp. 30--39.

\bibitem{li2018vuldeepecker}
Z.~Li, D.~Zou, S.~Xu, X.~Ou, H.~Jin, S.~Wang, Z.~Deng, and Y.~Zhong,
  ``{VulDeePecker: A Deep Learning-Based System for Vulnerability Detection},''
  in \emph{Proceedings of the 25th Annual Network and Distributed System
  Security Symposium (NDSS)}, 2018.

\bibitem{alqahtani2020webvuln}
S.~S. Alqahtani, E.~E. Eghan, and J.~Rilling, ``{Tracing Known Security
  Vulnerabilities in Software Repositories},'' \emph{Empirical Software
  Engineering}, vol.~25, no.~5, pp. 3867--3906, 2020.

\bibitem{sivanathan2019iot}
A.~Sivanathan, H.~H. Gharakheili, F.~Loi, A.~Radford, C.~Wijenayake,
  A.~Vishwanath, and V.~Sivaraman, ``{Classifying IoT Devices in Smart
  Environments Using Network Traffic Characteristics},'' in \emph{IEEE
  Transactions on Mobile Computing}, vol.~18, no.~8, 2019, pp. 1745--1759.

\bibitem{aafer2015android}
Y.~Aafer, W.~Du, and H.~Yin, ``{DroidAPIMiner: Mining API-Level Features for
  Robust Malware Detection in Android},'' in \emph{Security and Privacy in
  Communication Networks}, 2015, pp. 86--103.

\bibitem{li2026structural}
Z.~Li, B.~Ma, X.~Dai, M.~Xu, Y.~Zhang, B.~Yan, and K.~Li, ``Automating agent
  hijacking via structural template injection,'' \emph{arXiv preprint
  arXiv:2602.16958}, 2026.

\bibitem{maloyan2026breaking}
N.~Maloyan and D.~Namiot, ``Breaking the protocol: Security analysis of the
  model context protocol specification and prompt injection vulnerabilities in
  tool-integrated llm agents,'' \emph{arXiv preprint arXiv:2601.17549}, 2026.

\bibitem{radosevich2025mcp}
B.~Radosevich and J.~Halloran, ``Mcp safety audit: Llms with the model context
  protocol allow major security exploits,'' \emph{arXiv preprint
  arXiv:2504.03767}, 2025.

\bibitem{wang2025mcpguard}
B.~Wang, Z.~Liu, H.~Yu, A.~Yang, Y.~Huang, J.~Guo, H.~Cheng, H.~Li, and H.~Wu,
  ``Mcpguard: Automatically detecting vulnerabilities in mcp servers,''
  \emph{arXiv preprint arXiv:2510.23673}, 2025.

\bibitem{cve2024}
{MITRE Corporation}, ``{Common Vulnerabilities and Exposures (CVE)},''
  \url{https://cve.mitre.org}, 2024, accessed: 2026-02-27.

\bibitem{anthropic2025security}
{Anthropic}, ``{MCP Security Advisories},''
  \url{https://github.com/anthropics/mcp/security}, 2025.

\bibitem{jfrog2025mcp}
{JFrog Security Research}, ``{MCP Protocol Security Analysis},''
  \url{https://jfrog.com/blog/security-research-mcp}, 2025.

\bibitem{trend2025mcp}
{Trend Micro}, ``{Securing AI Interactions: MCP Vulnerability Report},''
  \url{https://www.trendmicro.com/en_us/research/25/mcp-security.html}, 2025.

\bibitem{cvss2023stats}
{FIRST CVSS SIG}, ``{CVSS Vulnerability Severity Statistics 2023},''
  \url{https://www.first.org/cvss/statistics}, 2023.

\bibitem{CVE-2025-68144}
{MITRE}, ``{CVE-2025-68144: Anthropic MCP Git Server git\_diff/git\_checkout
  Argument Injection Vulnerability},''
  \url{https://cve.mitre.org/cgi-bin/cvename.cgi?name=CVE-2025-68144}, 2025,
  cVSS v3: 8.1, CVSS v4: 6.4; Fixed in mcp-server-git 2025.12.18.

\bibitem{CVE-2025-68145}
------, ``{CVE-2025-68145: Anthropic MCP Git Server --repository Flag Path
  Validation Bypass},''
  \url{https://cve.mitre.org/cgi-bin/cvename.cgi?name=CVE-2025-68145}, 2025,
  cVSS v3: 7.1, CVSS v4: 6.3; Fixed in mcp-server-git 2025.12.18.

\bibitem{CVE-2026-22785}
------, ``{CVE-2026-22785},''
  \url{https://cve.mitre.org/cgi-bin/cvename.cgi?name=CVE-2026-22785}, 2026,
  unpublished/Under Review CVE Record.

\bibitem{CVE-2026-23523}
------, ``{CVE-2026-23523},''
  \url{https://cve.mitre.org/cgi-bin/cvename.cgi?name=CVE-2026-23523}, 2026,
  unpublished/Under Review CVE Record.

\bibitem{CVE-2025-66414}
------, ``{CVE-2025-66414},''
  \url{https://cve.mitre.org/cgi-bin/cvename.cgi?name=CVE-2025-66414}, 2025,
  cVSS v3: 8.1 (High Severity).

\bibitem{CVE-2026-0621}
------, ``{CVE-2026-0621},''
  \url{https://cve.mitre.org/cgi-bin/cvename.cgi?name=CVE-2026-0621}, 2026,
  published Jan 2026; High Severity.

\end{thebibliography}

\end{document}